\documentclass[%
 aip,
 jmp,%
 amsmath,amssymb,
reprint,%
]{revtex4}

\usepackage{graphicx}
\usepackage{dcolumn}
\usepackage{bm}

\begin{document}



\title[Fernandes-Santos]{Measurement of $p\!-\!n$-junction diode behavior under large signal and high frequency}
\thanks{One of the authors, EJPS, thanks CNPq for support.}

\author{Maria Augusta R. B. L. Fernandes}
\affiliation{
Laboratory for Devices and Nanostructures,\\
Engineering at Nanometer Scale Group,\\
Universidade Federal de Pernambuco, Recife-PE, Brasil.
}%

\author{Edval J. P. Santos}%
 \email{edval@ee.ufpe.br / e.santos@expressmail.dk.}
\affiliation{
Laboratory for Devices and Nanostructures,\\
Engineering at Nanometer Scale Group,\\
Universidade Federal de Pernambuco, Recife-PE, Brasil.
}%

\date{\today}

\begin{abstract}
Measurements of diode dynamic conductance and dynamic capacitance for frequencies up to $10 \times \tau_{p,n}^{-1}$, and voltage amplitude level up tp 100 mV was carried out with a precision impedancemeter.  The results were compared with the theoretical expressions obtained with the spectral approach to the charge carrier transport in $p\!-\!n$-junctions.  This experimental confirmation is of practical interest, as one can use the theory to extract device parameters, such as: relaxation time $\tau_{p,n}$, and junction injection coefficient.  These experiments were carried to test the extension of the conventional $p\!-\!n$-junction theory.
\end{abstract}



\maketitle

\section{Introduction}
\label{sec:Introduction}

Discrete $p\!-\!n$-junction diodes are fabricated with different techniques, namely: grown junction, alloy type or fused junction, diffused junction, epitaxial grown or planar diffused, and point contact~\cite{Streetman}.  The diode electrical behavior was explained by W. Shockley~\cite{Schokley1949}. His theory became the standard to describe the charge carrier transport in the  $p\!-\!n$-junction diode.  However, it was limited to small signals and low frequencies. This classical theory failed to explain the behavior of the $p\!-\!n$-junction diode when submitted to large signals at high frequencies.  In his book, S. M. Sze~\cite{Sze} gives an extension of Shockley's theory for high-frequencies, but only for small signals.  For large signals and low frequencies, it is well known that the current can be expressed as a Fourier series with coefficients proportional to Bessel functions, as shown by R. A. Schaefer~\cite{Schaefer1971,Pederson}.

A theory for the $p\!-\!n$-junction diode under large signals at high frequencies was missing. In a series of papers A.~A. Barybin and E.~J.~P. Santos have presented the unified theory of carrier transport in $p\!-\!n$-junction under large signals and high frequencies.  For the non-uniform junction case, to obtain the analytical expressions, the transverse averag\-ing technique was employed to reduce the three-dimensional charge carrier transport equations into the quasi-one-dimensional ones~\cite{SB2002,BS2007a,BS2007b,BS2011}.  Table~\ref{tab:pnjunctiontheories} is a summary of the many theories.

\begin{table}[h]
\begin{center}
\caption{PN theories.}
\begin{tabular}{|c|c|c|}
\hline
               & Small signal & Large signal \\
\hline\hline
               &             &                \\
Low frequency  & W. Schokley & R. A. Schaefer \\
               &             &                \\
\hline
               &             &                \\
High frequency & S. M. Sze      & A. A. Barybin \& \\
               &                & E. J. P. Santos\\
\hline
\end{tabular}
\end{center}
\label{tab:pnjunctiontheories}
\end{table}

The quasi-one-dimensional approximation is actually a good approximation for discrete diodes, as  for all common discrete $p\!-\!n$-junction diode fabrication techniques, an approximate axial symmetry can be assumed.  As a result of this approximation,  useful analytic expressions are obtained for the dynamic conductance, and dynamic (diffusion) capacitance as a function of the signal amplitude, frequency, and the cross section non-uniformity.

The purpose of this paper is to measure the {\it dynamic conductance\/} $G_d$ and the {\it dynamic\/} ({\it diffusion\/}) {\it capacitance\/} $C_d$ of real devices, and use the theoretical expression, with appropriate constants, to fit the experimental data to validate the theory.

The paper is divided into five sections, this introduction is the first.  Next, the basic expressions from the theory are presented. Thirdly, the measurements. Then, discussion, and finally the conclusions.

\section{Theory revisited}

\begin{figure}[t]
\centering
\includegraphics[scale=0.3, angle=0]{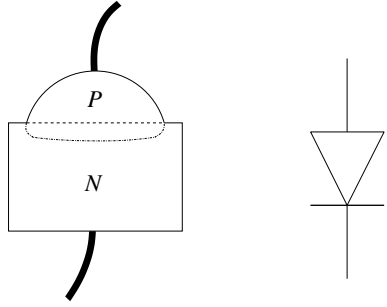}
\caption{Discrete $p\!-\!n$-junction diode can be fabricated with different techniques, such as: grown junction, alloy type or fused junction, diffused junction, epitaxial grown or planar diffused, and point contact.  For all such techniques, an approximate axial symmetry can be assumed.}
\label{diode}
\end{figure}

The $p\!-\!n$-junction diode is formed by doping a semiconductor material to create a p-type region next to a n-type region as shown in Figure~\ref{diode}.  For most applications, a diode is used under a time-varying signal, which is frequently of the form $V_{\sim} \cos(\omega t)$.

To estimate  $p\!-\!n$-junction diode behavior when operating under such signals, it is necessary to calculate the admittance, $Y_d= G_d + j\omega C_d$, in which the {\it dynamic conductance\/} $G_d$ and the {\it dynamic\/} ({\it diffusion\/}) {\it capacitance\/} $C_d$ are defined as customary~\cite{Sze}.  It has been demonstrated that for the uniform cross-section diode, a unified theory for large signal and high frequency yields the following expressions~\cite{BS2007b}.

\begin{equation}
{G_d(\omega,V_{\sim})\over G_{d0}(V_0)} =
{ \sqrt{1+\sqrt{1+ \omega^2\tau_p^2}}\over\sqrt{2}}\;
{\,I_1(\beta V_{\sim})\over \beta V_{\sim}/2} \,,
\label{eq:01}
\end{equation}
\begin{equation}
{C_d(\omega,V_{\sim})\over C_{d0}(V_0)} =
{ \sqrt{2}\over\sqrt{1+\sqrt{1+ \omega^2\tau_p^2}}}\;
{\,I_1(\beta V_{\sim})\over \beta V_{\sim}/2} \,,
\label{eq:02}
\end{equation}
in which, $I_1(\beta V_{\sim})$ is the modified Bessel function~\cite{GR1980}, and $\beta\equiv {q\over \kappa T}$. 

These expressions take into account the signal amplitude influence owing to the factor $I_1(\beta V_\sim)/(\beta V_\sim /2)$.  The frequency dependence is included in the  $\omega\tau_p$ term.

At low-frequency values ($\omega\tau_p\!\ll 1$), the dynamic conductance, $G_d$, and the  dynamic diffusion capacitance, $C_d$, satisfy the following equations.
\[
{G_d(0,V_{\sim})\over G_{d0}(V_0)} =
{C_d(0,V_{\sim})\over C_{d0}(V_0)} =
{I_1(\beta V_{\sim})\over \beta V_{\sim}/2}.
\]

These expressions have been generalized for the non-uniform cross-section case~\cite{BS2011}. In this generalized case, new functions $F_G^\alpha(\omega)$ and  $F_C^\alpha(\omega)$ are introduced.

\begin{equation}
{G_d(\omega,V_{\sim})\over G_{d0}(V_0)} \simeq F_G^\alpha(\omega)\,
{I_1(\beta V_{\sim})\over \beta V_{\sim}/2}\,,  \qquad
\label{eq:03}
\end{equation}
\begin{equation}
{C_d(\omega,V_{\sim})\over C_{d0}(V_0)} \simeq F_C^\alpha(\omega)\,
{I_1(\beta V_{\sim})\over \beta V_{\sim}/2}\,.  \qquad
\label{eq:04}
\end{equation}
in which, the frequency-dependent factors are as follows.

{\small
\begin{eqnarray}
F_G^\alpha(\omega) &=&
\bigl[ \alpha L_p + \nonumber \\
&&\frac{e^{2\alpha W_n}}{\sqrt2}\,\sqrt{1\!+\!(\alpha L_p)^2\!+
\sqrt{\bigr[1\!+\!(\alpha L_p)^2 \bigr]^2\!+(\omega\tau_p)^2}} \bigr]\,\,\,\,\,\,\,
\label{eq:05}\\
F_C^\alpha(\omega) &=&
\frac{\sqrt{2}\,\,e^{2\alpha W_n}}{\sqrt{1\!+\!(\alpha L_p)^2\!+
\sqrt{\bigr[1\!+\!(\alpha L_p)^2 \bigr]^2\!+(\omega\tau_p)^2}}}
\label{eq:06}
\end{eqnarray}
}in which, the cross section $S(z)= S_0\, \exp(2\alpha z)$, $\alpha$ is the parameter which affects the cross section change, $L_p= \sqrt{D_p\tau_p}$ is the diffusion length.

For the uniform $p\!-\!n$-junction, $\alpha =0$, Eqs.~(\ref{eq:05}) and (\ref{eq:06}) reduce to the frequency-dependent factors in Eqs.~(\ref{eq:01}) and (\ref{eq:02}).
\begin{equation}
F_G^\alpha(\omega)\,\stackrel{\alpha\to 0\vphantom{j}}{\longrightarrow}\,
{\sqrt{1+\sqrt{1+(\omega\tau_p)^2}}\over\sqrt{2}}
\label{eq:4.81}
\end{equation}
\begin{equation}
F_C^\alpha(\omega)\,\stackrel{\alpha\to 0\vphantom{j}}{\longrightarrow}\,
{\sqrt{2}\over\sqrt{1+\sqrt{1+(\omega\tau_p)^2}}}
\label{eq:4.82}
\end{equation}

\section{Measurements}

Measurements were carried out with the Agilent 4294A  precision impedancemeter.  Diode conductance and capacitance are measured from 100 Hz up to 100 kHz, and voltage amplitude at 10 mV, 20 mV, 50 mV, and 100 mV levels.  The high frequency is selected such as it is much higher than $\tau_{p,n}^{-1}$. The system calibration plane is located where the cable connects the diode mounting board. Hence, there is no cable capacitance to be taken into consideration, only the mounting board parasitic capacitance.  

\begin{figure}[t]
\centering
\includegraphics[scale=0.4, angle=0]{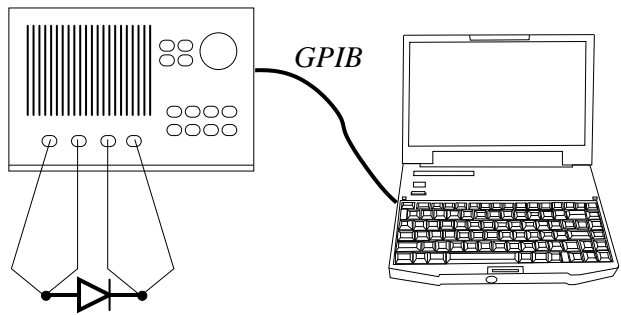}
\caption{Experimental setup with the Agilent 4294A precision impedancemeter Agilent connected to a computer over GPIB.}
\label{setup}
\end{figure}

Data is collected by a computer over a GPIB connection. To test the parameter extraction procedure, measurements were carried out on 1N4148 diodes.  Curve fitting is carried out with MATLAB. The experimental setup is presented in Figure~\ref{setup}.

\section{Discussion}

\begin{figure}[t]
\centering
\includegraphics[scale=0.4, angle=0]{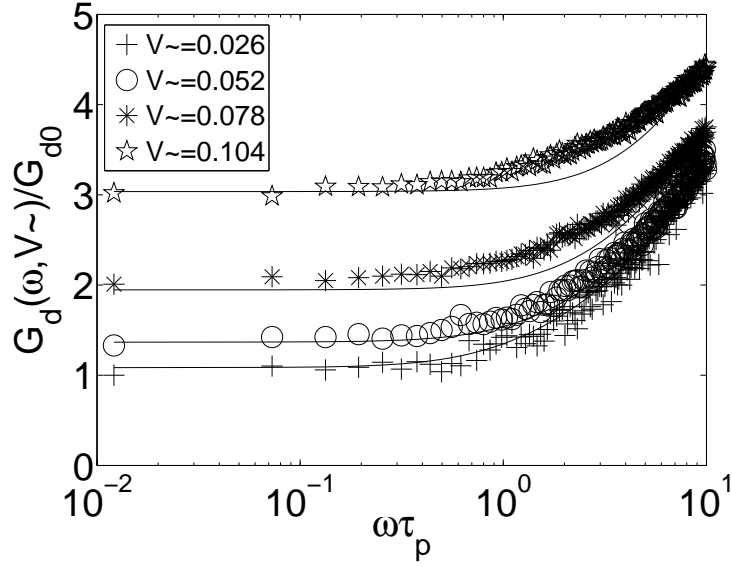}
\caption{Measured dynamic conductance as a function of frequency and voltage level.  To achieve the fitting, a different relaxation is obtained for each curve.}
\label{1n4148Gind}
\end{figure}

After collecting the conductance and the capacitance for 1N4148 junction diodes, one can proceed to extract the diode parameters. The first fitting was performed in MATLAB considering the uniform cross-section theory. The result is presented in Figure~\ref{1n4148Gind}. 

For this fit, Eq.~\ref{eq:01} and Eq.~\ref{eq:02} are used. A good fit is obtained, but the relaxation time changes with the applied voltage, as can be seen in Table~\ref{diode1N4148}. For these measurements $\beta_n = 31.1864$.  Assuming $\kappa T/q = 25.85\,mV$, ($\beta= 38.64$), one can calculate the junction injection coefficient, $n = \beta_n/\beta= 1.2$.

 \begin{table}[htbp]
    \centering
    \caption{Voltage dependent relaxation time for the 1N4148.}
    \begin{tabular}{c|c}
    \hline
    voltage level & relaxation time,  \\
     (mV)         & $\tau_{p,n} $ (s)  \\
    \hline\hline
    104  & $0.63\times 10^{-5}$  \\
     78  & $1.27\times 10^{-5}$ \\
     52  & $2.25\times 10^{-5}$ \\
     26  & $3.54\times 10^{-5}$ \\
    \hline
    \end{tabular}
    \label{diode1N4148}
    \end{table}

\begin{equation}
C_{d0}(V_0)= G_{d0}(V_0)\frac{\tau_p}{2}
\label{eq:09}
\end{equation}

To get a good fit for the capacitance curves, one has to subtract the parasitic capacitance, which for the constructed experimental setup is $C_{\mbox{parasitic}}= 0.65 pF$.  This parasitic capacitance is related to the mounting structure in the printed circuit board, and not related to the cable, as the calibration plane includes the cable. The value of the parasitic capacitance is extracted by comparing data and theory.

\begin{figure}[h]
\centering
\includegraphics[scale=0.4, angle=0]{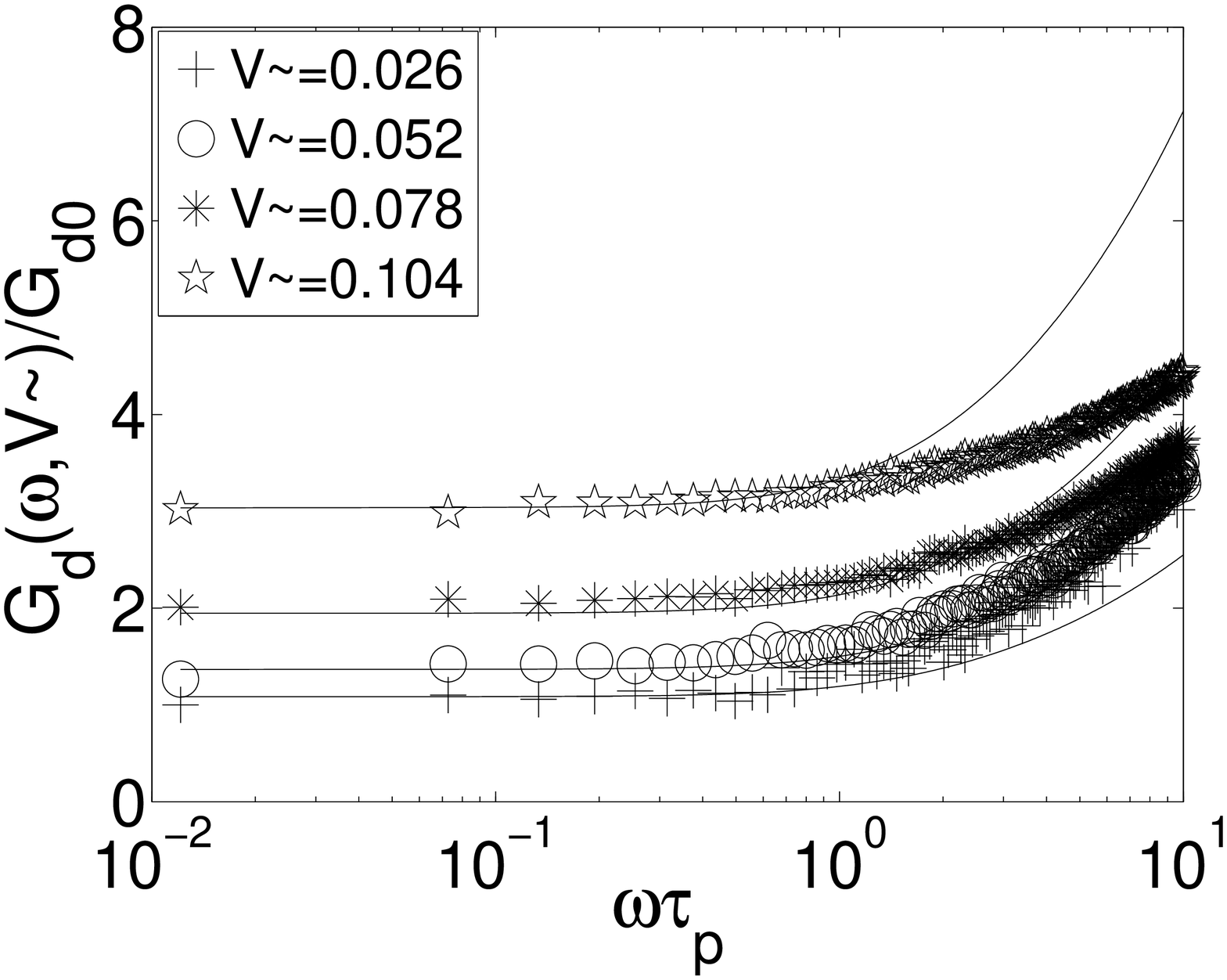}
\includegraphics[scale=0.4, angle=0]{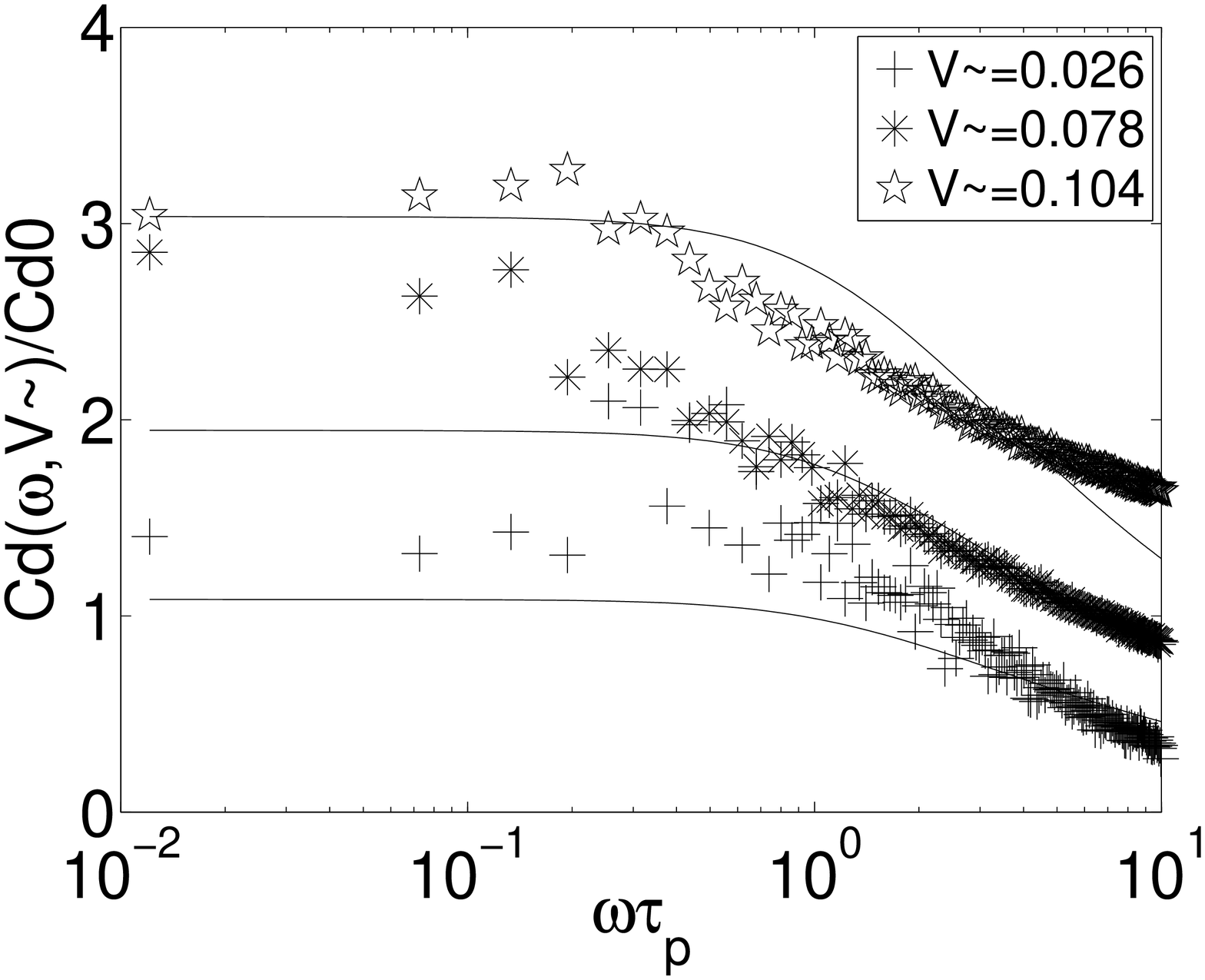}
\caption{Top: Dynamic conductance fitting using an average relaxation time.
Bottom: Dynamic capacitance fitting using an average relaxation time.}
\label{1n4148Gmed}
\end{figure}

\begin{figure}[h]
\centering
\includegraphics[scale=0.4, angle=0]{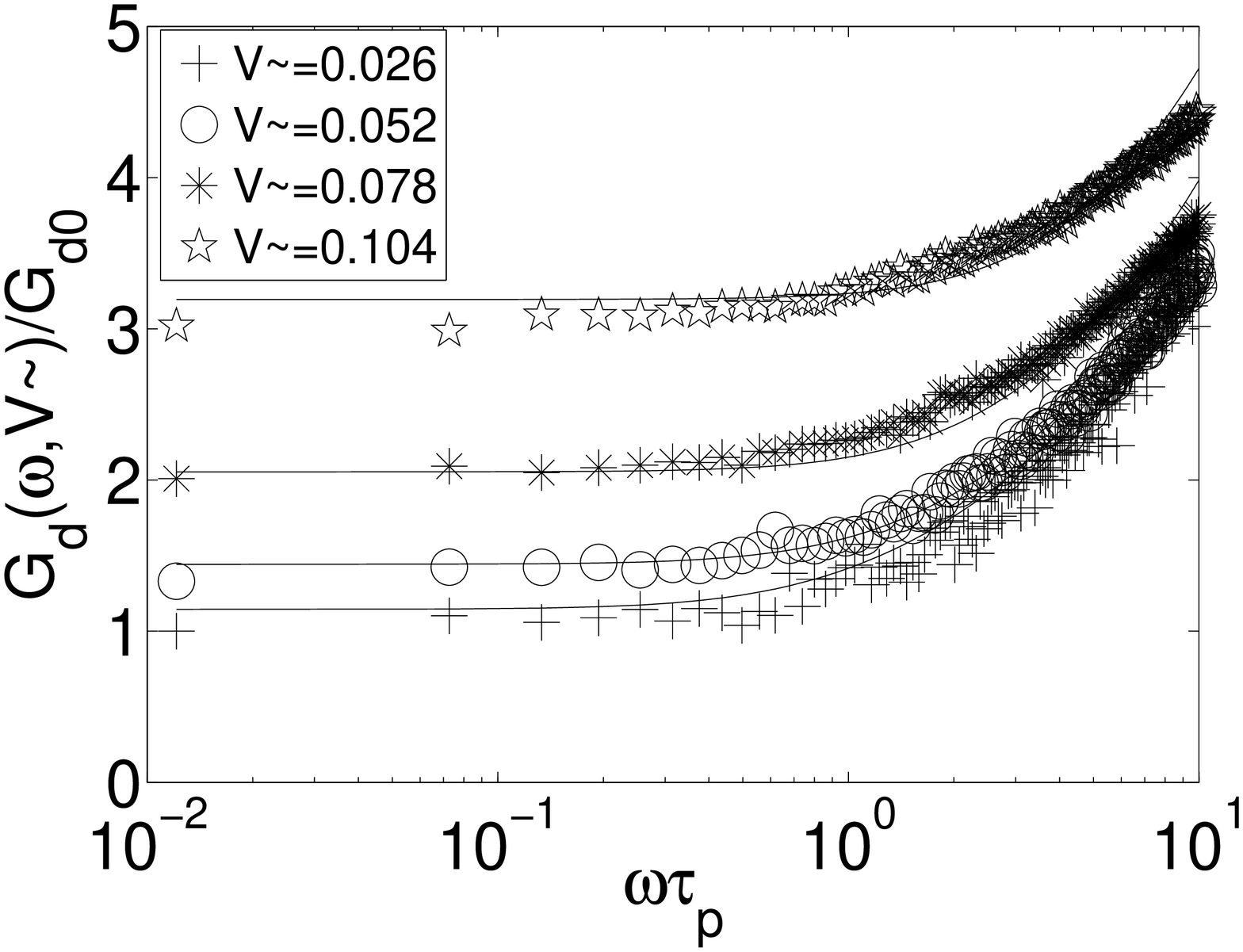}
\includegraphics[scale=0.4, angle=0]{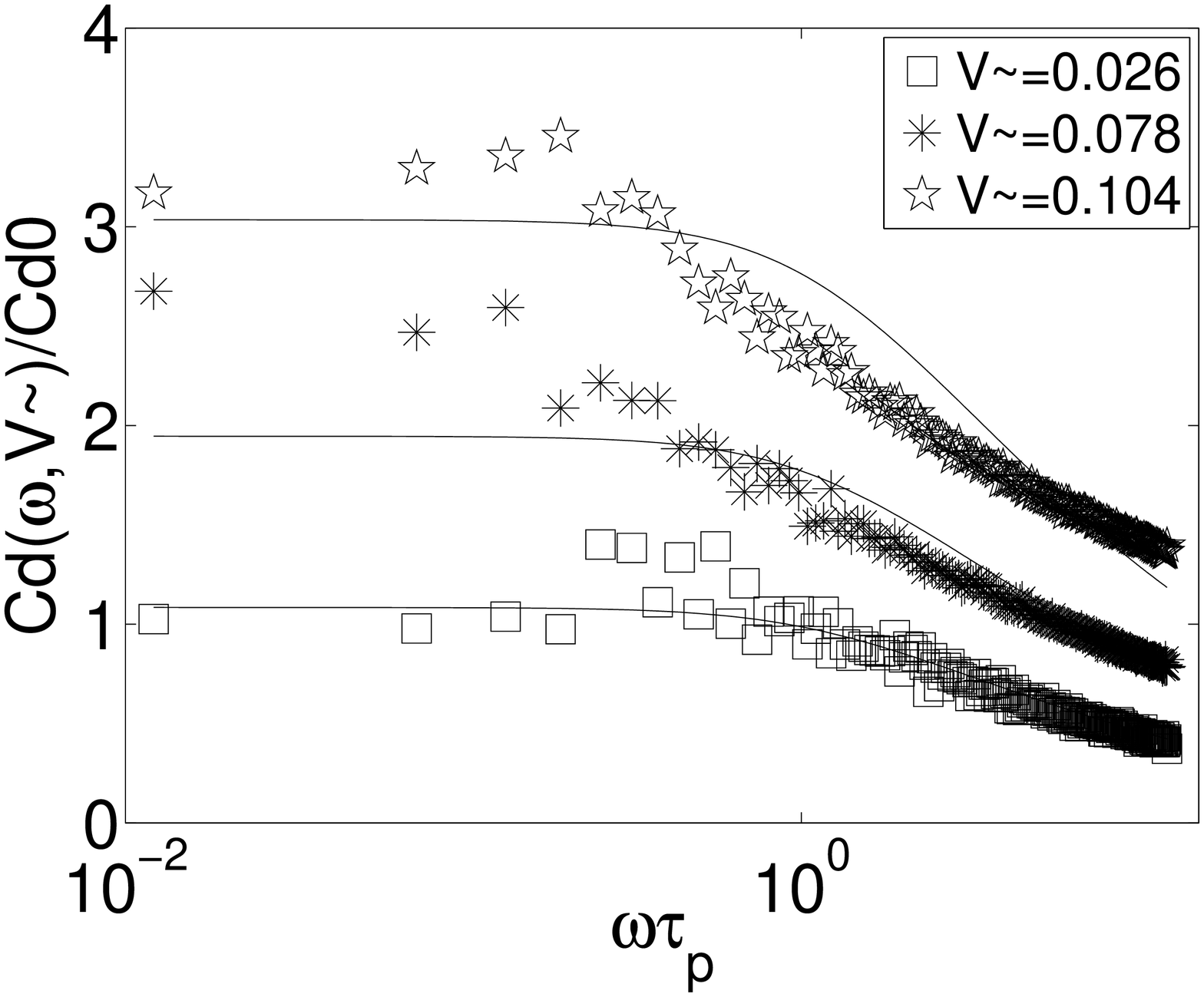}
\caption{Top: Dynamic conductance fitting using the variable cross-section theory.  Bottom: Dynamic capacitance fitting.}
\label{1n4148Gajuste2}
\end{figure}

As the relaxation time should not change with voltage level, an average value of the extracted relaxation times was calculated and use to fit the data.  It is observed that the curve fitting gets worse, as presented in Figure~\ref{1n4148Gmed}.  For the dynamic capacitance, the result is also presented in Figure~\ref{1n4148Gmed}. 

This voltage dependent relaxation time may indicate that for the 1N4148, the uniform junction theory is not adequate.  The next step was to fit the data by applying the nonuniform junction theory.  Example of non-idealities are non-uniform junction doping or cross-section area variation, affecting the field distribution inside the device~\cite{BS2011}.  Considering the extended version of the theory, new plots were prepared, with the parameters presented in Table~\ref{diode1N4148var}. The same value for the parasitic capacitance, $C_{\mbox{parasitic}}= 0.65\, pF$, was subtracted.  The resulting curve fitting is presented in Figure~\ref{1n4148Gajuste2}, which display a much better fit, as compared to Figure~\ref{1n4148Gmed}.

 \begin{table}[htbp]
    \centering
    \caption{Variable cross-section parameters for the 1N4148, with constant relaxation time.}
    \begin{tabular}{l|c}
    \hline
     Parameter & Value  \\
    \hline\hline
    Carrier lifetime ($\tau_p$) & $1.93\times 10^{-5}$ s\\
    Diffusion length ($L_p$)  & $1.5\times 10^{-6}$ m  \\
    Depletion layer interface ($W_n$)  & $1\times 10^{-6}$ m\\
    Cross-section variation factor ($\alpha$) & $1.555 \times 10^4$ m$^{-1}$\\
    \hline
    \end{tabular}
    \label{diode1N4148var}
    \end{table}

\section{Conclusion}

For the first time, measurements of diode conductance and capacitance are presented to test the spectral approach theory to the charge carrier transport in $p\!-\!n$-junctions. The experimental results are nicely explained by the extended $p\!-\!n$-junction theory.  From the measurement results performed on 1N4148 diodes, a voltage dependent relaxation times are extracted, if the  uniform junction theory is used to fit the data. However, after applying the non-uniform junction theory, a better fit is achieved with constant relaxation times.  The experimental results suggest that the 1N4148 diode has a non-uniform junction, which is not surprising.  Hence, the theory can be applied for diode parameter extraction.

The measurements were carried out for frequencies up to 100 kHz, and voltage amplitude level up to 100 mV,  with a precision impedancemeter Agilent 4294A.  Further evaluation is in progress to extract better values for the various parameters.  The experiment will be performed with other diodes, and for the bipolar transistor base-emitter junction.

\section*{Acknowledgment}

This paper is offered in memoriam of Dr. A. A. Barybin, who was a visiting scientist at the Laboratory for Devices and Nanostructures at UFPE, Recife, Brazil.  He died in 2011.

\bibliographystyle{model1-num-names}
\bibliography{<your-bib-database>}

\begin{thebibliography}{1}

\bibitem{Streetman}
B. G. Streetman, ``Solid state electronic devices'', 2nd. Ed., Prentice-Hall Inc., Englewood Cliffs, N.J. (1980).

\bibitem{Schokley1949}
W. Shockley, ``The Theory of p-n Junctions in Semiconductors and p-n Junction Transistors'', The Bell System Technical Journal, {\bf 28}, 435 - 489 (1949).
%
\bibitem{Sze}
S. M. Sze, ``Physics of Semiconductor Devices'', Wiley, New York (1969),  2nd ed. (1981),  3rd ed., with K. K. Ng (2006).
%
\bibitem{Schaefer1971}
R. A. Schaefer, ``Production of Harmonics and Distortion in p-n Junctions'',
J. Audio Eng. Soc., {\bf 19}, 759 - 769 (1971). 
%
\bibitem{Pederson}
D. O. Pederson, K. Mayaram, ``Analog integrated circuits for communication'', Kluwer Academic Publishers (1991).
%
\bibitem{SB2002} E. J. P. Santos and A. A. Barybin, ``Large-signal
dynamic admittance of $p\!-\!n$-junctions'',  Proc. of the
XVII Int. Symp. on Microelectronics Technology and Devices,
Electrochem. Soc. Proc., PV2002-8 (2002) 237-243.
``Novel Results on the Large-Signal Dynamic Admittance of $p-n$-Junctions'',
cond-mat/0204620.
%
\bibitem{BS2007a} A. A. Barybin and E. J. P. Santos, `` Transverse averaging
technique for the depletion capacitance of nonuniform PN-junctions'',
Semicond. Sci. Technol. {\bf 22} (2007) 312-319.
%
\bibitem{BS2007b} A. A. Barybin and E. J. P. Santos, ``Unified Approach to the
Large-Signal and High-Frequency Theory of PN-Junctions'',
Semicond. Sci. Technol. {\bf 22} (2007) 1225-1231.
%
\bibitem{BS2011}
A. A. Barybin and E. J. P. Santos, ``Large-signal and high-frequency analysis of nonuniformly doped or shaped pn-junction diodes'',
J. Appl. Phys. {\bf 109}, 114510 (2011). 
%
\bibitem{GR1980} I. S. Gradshteyn and I. M. Ryzhik, {\em Tables of Integrals,
Series, and Products.\/} New York: Academic Press, 1980.
%
\end{thebibliography}

\vfil

\end{document}